\documentclass[aps,physrev,reprint]{revtex4-2}

\usepackage{graphicx,amsmath}
\usepackage{braket}
\usepackage{dcolumn}
\usepackage{bm}

\begin{document}


\title{Connection Between Classical and Quantum Descriptions of Spin Waves Using Quantum Circuits}



\author{Daniel D. Stancil}
\affiliation{Department of Electrical and Computer Engineering, North Carolina State University, Raleigh, NC 27695 USA}
\email{ddstancil@ncsu.edu}

\author{Bojko N. Bakalov}
\affiliation{Department of Mathematics, North Carolina State University, Raleigh, NC 27695 USA}

\author{Gregory T. Byrd}
\affiliation{Department of Electrical and Computer Engineering, North Carolina State University, Raleigh, NC 27695 USA }


\date{\today}

\begin{abstract}
A quantum computing circuit is presented that approximates a single spin wave quantum on a linear chain of spin 1/2 particles described by a Heisenberg Hamiltonian. The circuit is a product state where each qubit represents a spin. The spin wave motion is represented by opening the cone angle using $Y$ rotations and then adding progressive $Z$ rotations along the chain to represent wave propagation. We show analytically that 
this product state yields the correct dispersion relation in the limit of an unbounded chain. 
This surprising observation is confirmed using both a simulator and various quantum processors. The quantum circuit calculation 
leads to insight into the connection between classical and quantum descriptions of spin waves, and may also be useful for characterizing the error in quantum processors.
\end{abstract}


\maketitle


\section{Introduction}

The classical picture of spin waves consists of spins on each lattice site precessing like a spinning top, but with a linear phase variation with position. In contrast, the quantum mechanical picture of the lowest-energy spin wave quantum consists of a single flipped spin, with equal probability of being on each lattice site, and with a linear phase of the probability amplitudes with position.

Although to truly make the quantum-to-classical transition requires large numbers of spin wave quanta, a connection can be made with quantum circuits that gives some insight.

\begin{figure}
    \centering
    \includegraphics[width=\linewidth]{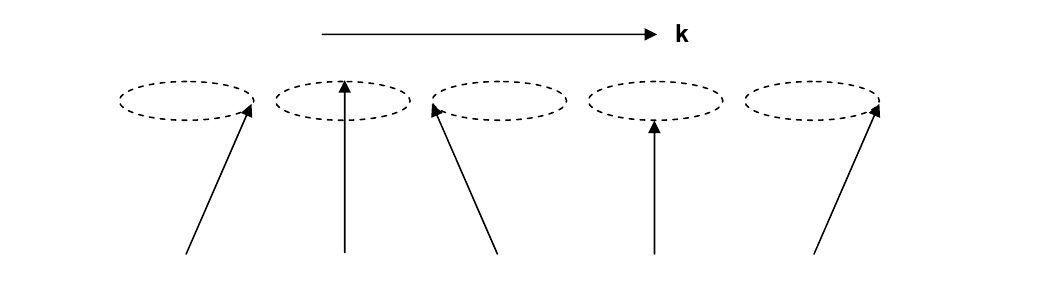}
    \caption{Classical depiction of a spin wave (Reproduced from \cite{stancil_spinwave_2009}with permission).}
    \label{fig:spinwave}
\end{figure}

In particular, consider a linear chain of spins, each represented by a qubit. Motivated by the classical picture, we might consider an attempt to represent a spin wave by rotating each spin along the Bloch sphere by a small angle representing the classical ``cone angle," and then applying increasing $Z$ rotations along the chain to represent the progressive phase associated with the wave, as shown in Figure \ref{fig:spinwave}. The states would then be precessing about the $Z$ direction with a phase that increases linearly with position, reminiscent of our classical picture. In this work, we explore this construction and its connection with the quantum spin wave state.

\section{Semi-Classical Description}
The semi-classical description of spin waves has been discussed by numerous authors. A useful overview for our purposes was given by Morgenthaler \cite{Morgenthaler_overview_1988}, in which the number of flipped spins per unit volume $N_f/V$ is given by
\begin{equation}
\frac{N_f}{V} = 
\frac{1}{2}\mu_0 \frac{\langle |m|^2 \rangle}{\hbar \omega_M},
\label{eq:NfperV}
\end{equation}
where $|m|$ is the magnitude of the small-signal magnetization, $\omega_M = |\gamma|\mu_0 M_S$, $|\gamma|= e/m_e$ with $m_e$ and $e$ being the electron mass and charge magnitude respectively, $\mu_0$ is the permeability of free space, and $M_S$ is the saturation magnetization of the sample. If the precession of $\mathbf{m}$ is circular, then the number of flipped spins is equal to the number of magnons and $|m|^2 = m_0^2$ \cite{Morgenthaler_overview_1988}. If $\theta$ is the angle that the magnetization makes to the $z$ (equilibrium) axis (i.e., the ``cone angle"), then $m_0 = M_S\sin{\theta}$, where $M_S = gNS\mu_B/V$, $g$ is the Land\'{e} $g$ factor ($g=2$ for spin), $N$ is the total number of sites with spin, $S$ is the spin per site, and $\mu_B=e\hbar/2m_e$ is the Bohr magneton. 

To explore the case of a single flipped spin/magnon, we set $N_f=1$. Substituting the above expressions for $|m|$ and $\omega_M$ into Eq. (\ref{eq:NfperV}) and simplifying leads to
\begin{equation}
    \sin{\theta}=\sqrt{\frac{2}{NS}}.
    \label{eq:theta}
\end{equation}
For spin 1/2, this reduces to
\begin{equation}
    \sin{\theta} =  \frac{2}{\sqrt{N}},
    \label{eq:thetahalf}
\end{equation}
which for small angles ($N\gg1$) is approximated by
\begin{equation}
    \theta \approx \frac{2}{\sqrt{N}}.
    \label{eq:sstheta}
\end{equation}

\begin{figure}[tbh]
\centerline{\includegraphics[width=1.5in]{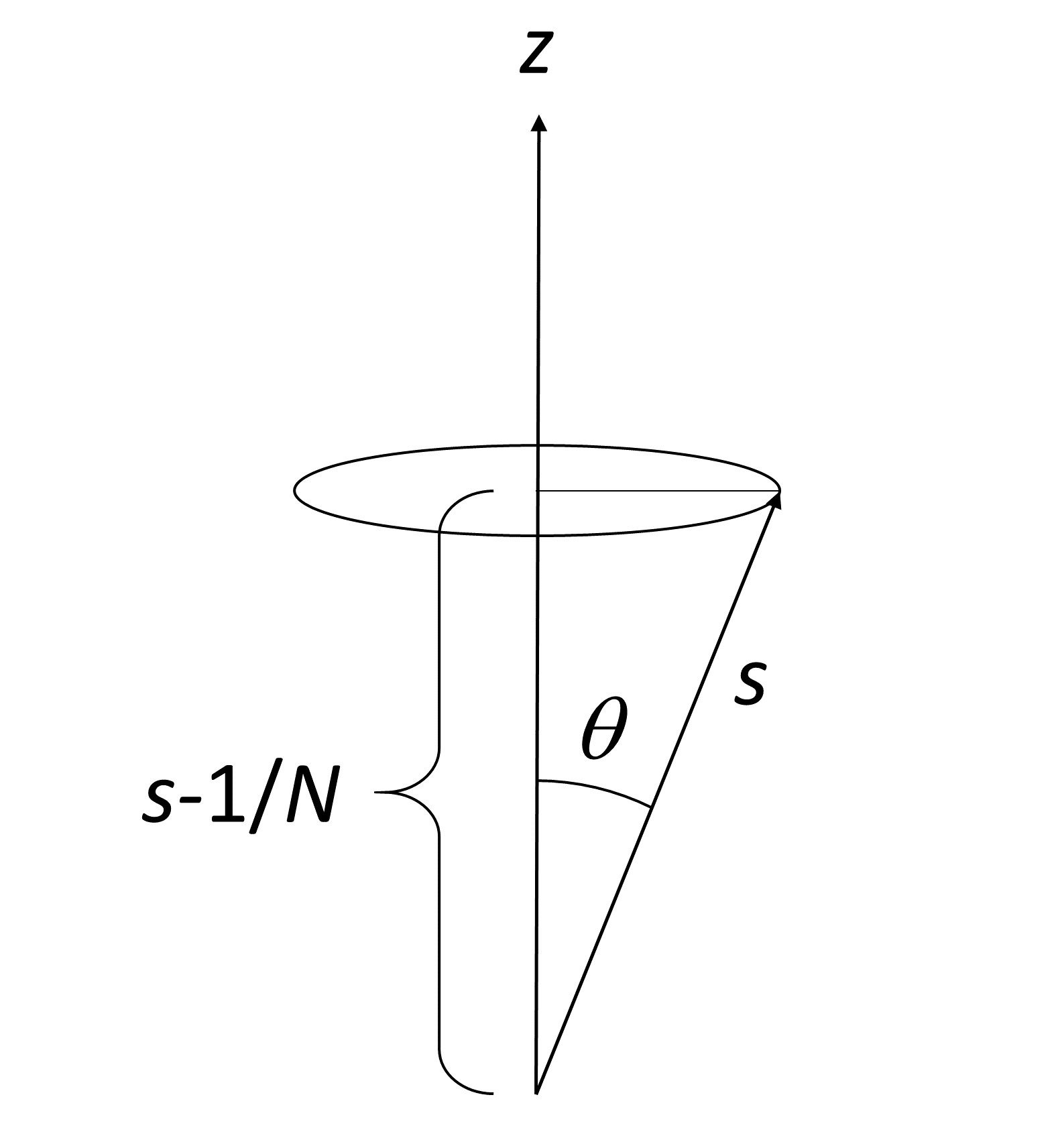}}
\caption{Classical vector of length $S$ precessing about the $z$-axis. (Reproduced from \cite{stancil_spinwave_2022}  with permission).}
\label{fg:coneangle}
\end{figure}

For an equivalent geometry-motivated approach, consider the precession of a classical vector of length $S$ precessing with a cone angle of $\theta\ll1$ as shown in Figure \ref{fg:coneangle}. If $S$ represents the total spin angular momentum, then flipping a single spin 1/2 component would reduce the total $z$ component of spin to $S-1$. If there is a single flipped spin on a lattice with $N$ sites, then the average reduction in the $z$ component of the spin on a given site will be $S-1/N$. It follows that
\begin{equation}
    \cos{\theta}=\frac{S-1/N}{S}=1-\frac{1}{NS}.
\end{equation}
Expanding the cosine for small angles again leads to Eq. (\ref{eq:sstheta}).

\section{Quantum Description}
The quantum state for a single spin wave quantum on a linear lattice can be written as \cite{Dyson_general_1956,ashcroft_1976,kittel_1991,stancil_spinwave_2009}
\begin{equation}
    \ket{\psi} = \sum_n \ket{n}\braket{n|\psi},
    \label{eq:quantstate}
\end{equation}
where $\ket{n}$ is the state with a single flipped spin on site $n$, and the probability amplitudes are
\begin{equation}
    \braket{n|\psi} = \frac{e^{ik_m x_n}}{\sqrt{N}},
    \label{eq:coeff}
\end{equation}
where $x_n=n a$ is the $n$th lattice site and the spacing between the sites is $a$. For periodic boundary conditions, $\braket{N|\psi}=\braket{0|\psi}$, and the wavenumber is quantized to the values $k_m = 2\pi m/Na,\quad m=0,\pm 1,\pm 2,\pm3,$ (assuming that $N$ is even for simplicity). Here, the $+/-$ signs correspond to propagation in the $+/-$ directions along the chain, respectively, assuming a time dependence of $\exp({-i\omega t})$.

As an example, a spin wave on a chain of 4 spin 1/2 sites with a wavelength equal to the length of the chain can be written
\begin{align}
    \ket{\psi}&=\frac{1}{2}\left(\ket{0001}+ e^{-ika}\ket{0010}+\nonumber\right. \\ 
    &\quad +\left. e^{-i2ka}\ket{0100}+ e^{-i3ka}\ket{1000}\right),
    \label{eq:quantstate4}
\end{align}
where $k=\pi/2a$.

\section{Circuit Construction of the Quantum State}
We now consider how to construct an appropriate quantum state using quantum computing circuits. One way to construct states such as (\ref{eq:quantstate4}) is using the mathematical approach described in \cite{Plesch_quantum_2011,Ranu_single_2022}.
However, this approach is computationally involved and does not lend itself readily to an intuitive understanding. Akhil et al. \cite{francis_2020} obtained the spin wave spectrum for two- and four-spin chains from the temporal correlation function. However, extensions of this approach require full Trotterization, and so are also computationally intensive.

Motivated by the semi-classical picture of precessing spins and a progressive phase shift with position, let us attempt to construct an analogous state using single qubit rotations.

Just as in the more mathematical approach mentioned above, we assume that each qubit represents a spin on a lattice. Generally speaking, the state of each single qubit is represented by a point on the Bloch sphere using the well-known expression
\begin{equation}
    \ket{\psi} = \cos{\frac{\theta}{2}}\ket{0}+e^{-i\phi}\sin{\frac{\theta}{2}}\ket{1}.
\end{equation}
It seems reasonable to consider $\theta$ to be analogous to the classical cone angle. A progressive phase $\phi_i$ would then represent the phase shift in (\ref{eq:coeff}).
If we consider a single flipped spin to be equally likely on any site in the chain, then the probability of a flipped spin on a given site is
\begin{equation}
    \sin^2{\frac{\theta}{2}} = \frac{1}{N},
\end{equation}
where $N$ is the number of spins in the chain. For $\theta\ll1$ ($N\gg1$), this reduces to Eq. (\ref{eq:sstheta}), in agreement with the semi-classical picture. We therefore proceed to construct a state where each qubit is rotated about the $Y$ axis by angle $\theta$, and then each qubit is rotated about the $Z$ axis by angle $nka$, where $n$ is the qubit number, $k$ is the wave number, and $a$ is the assumed physical separation of the spins. The $N$-qubit state ansatz then takes the form
\begin{align}
    \ket{\psi_A}&=\left(\cos{\frac{\theta}{2}}\ket{0}+\sin{\frac{\theta}{2}}\ket{1}\right)\nonumber\\
    &\quad \otimes \left(\cos{\frac{\theta}{2}}\ket{0}+e^{\pm ika}\sin{\frac{\theta}{2}}\ket{1}\right)\nonumber\\
    &\quad \otimes \left(\cos{\frac{\theta}{2}}\ket{0}+e^{\pm i2ka}\sin{\frac{\theta}{2}}\ket{1}\right) \otimes\cdots\nonumber \\
    &=\bigotimes_{n=0}^{N-1}\left(\cos{\frac{\theta}{2}}\ket{0}+e^{\pm inka}\sin{\frac{\theta}{2}}\ket{1}\right).
    \label{eq:ansatz}
\end{align}
Here the $\pm$ signs represent waves traveling in the minus and plus directions, respectively.

Unlike (\ref{eq:quantstate}), however, the resulting state is a superposition of all possible $N$-qubit basis states. The amplitudes of the states are proportional to
\begin{equation}
    \left(\cos{\frac{\theta}{2}}\right)^{N-m}\left(\sin{\frac{\theta}{2}}\right)^m,
    \label{eq:basisamp}
\end{equation}
where $m$ is the number of flipped spins in the basis state. Since we are interested in states with a single flipped spin, it is interesting to find the angle $\theta$ that maximizes the value of (\ref{eq:basisamp}) for $m=1$. Specifically, taking the derivative of (\ref{eq:basisamp}) with respect to $\theta$ for $m=1$ and setting it to zero gives
\begin{equation}
    \sin^2{\frac{\theta}{2}} = \frac{1}{N},
    \label{eq:optangle}
\end{equation}
which again reduces to (\ref{eq:sstheta}) for $N\gg1$.

\begin{figure}
    \centering
    \includegraphics[width=\linewidth]{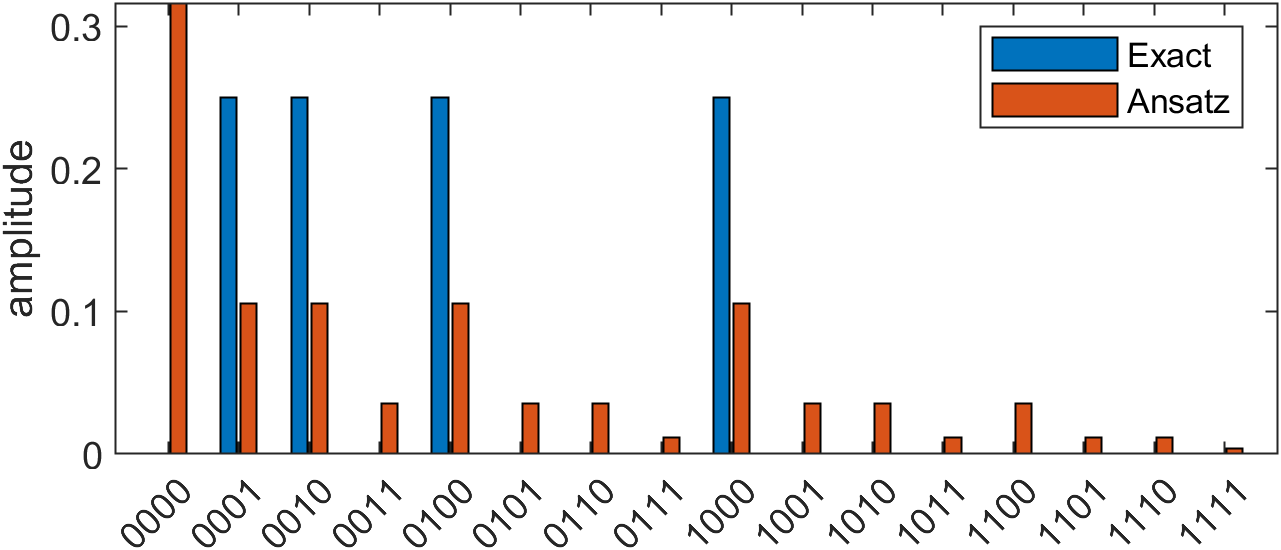}
    \caption{Probability amplitudes  of the basis states for both the exact wavefunction and the Ansatz.}   \label{fig:wavefunctions}
\end{figure}

To compare the ansatz using the optimum angle given by Eq. (\ref{eq:optangle}) with the exact wavefunction, the probabilities of the basis states for $N=4$ are shown in Fig. \ref{fig:wavefunctions}. Apart from the large ground state component, the basis states having a single flipped spin have the largest amplitudes. It is interesting to note that for large $N$, the amplitude of the ground state component has the limiting value of (calculated with the assistance of Wolfram$|$Alpha \cite{wolframalpha2025}) 
\begin{equation}
    \lim_{N\rightarrow\infty} \left(\cos\frac{\theta}{2}\right)^N = \frac{1}{\sqrt{e}}.
\end{equation}

\begin{figure}
    \centering
    \includegraphics[width=0.4\textwidth]{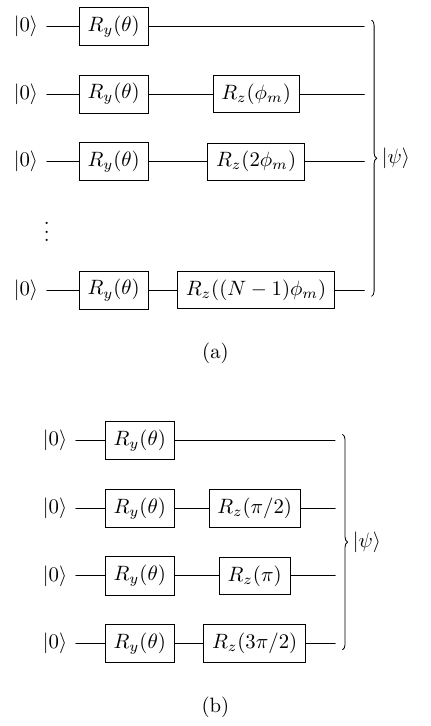}
    \caption{Circuits to construct quantum states containing a spin wave quantum on a linear spin chain. (a) $N$ spin linear chain, $\phi_m=2\pi m/N$. (b) 4 spin chain with the wavelength equal to the length of the chain. }
    \label{fig:quantumcircuits}
\end{figure}

A circuit for generating the ansatz state for an $N$-spin chain is shown in Fig. \ref{fig:quantumcircuits}(a), while a specific example for 4 spins with wavelength equal to the chain length is shown in Fig. \ref{fig:quantumcircuits}(b).

\section{Dispersion Relation Calculation}
The Hamiltonian for the linear spin chain with nearest-neighbor interactions is given by
\begin{equation}
H=-\frac{J}{2}\sum_{i=0}^{N-1}\boldsymbol{\sigma}_i\cdot\boldsymbol{\sigma}_{(i+1)\text{mod}\,{N}},\label{eq:Heisenberg_ham}
\end{equation}
where $J$ is the exchange coupling constant, $\boldsymbol{\sigma}_i$ is the Pauli vector for site $i$, and the mod $N$ index calculation imposes periodic boundary conditions.


The theoretical dispersion curve (energy vs. wavenumber) for a chain with $N$ spins is given by \cite{stancil_spinwave_2009}:
\begin{equation}
    \frac{\hbar\omega}{J/2} +N = 4(1-\cos{k_m a}).
    \label{eq:theory}
\end{equation}
In the limit $N\rightarrow\infty$ the wavenumber $k_m$ becomes a continuous variable.

An estimate of $\hbar\omega$ can be obtained by taking the expectation value of the
Hamiltonian (\ref{eq:Heisenberg_ham}) using the ansatz (\ref{eq:ansatz}).
A direct calculation of this expectation value (see Appendix \ref{app:A}) leads to
\begin{equation} 
    \lim_{N\rightarrow\infty}\left(\frac{2}{J}\braket{\psi_A|H|\psi_A} +N\right) = 4(1-\cos{ka}).
    \label{eq:limit}
\end{equation}
This is particularly surprising, since (\ref{eq:quantstate}), (\ref{eq:quantstate4}) are entangled states, while (\ref{eq:ansatz}) is explicitly a product state. 

\begin{figure}
    \centering
    \includegraphics[width=0.75\linewidth]{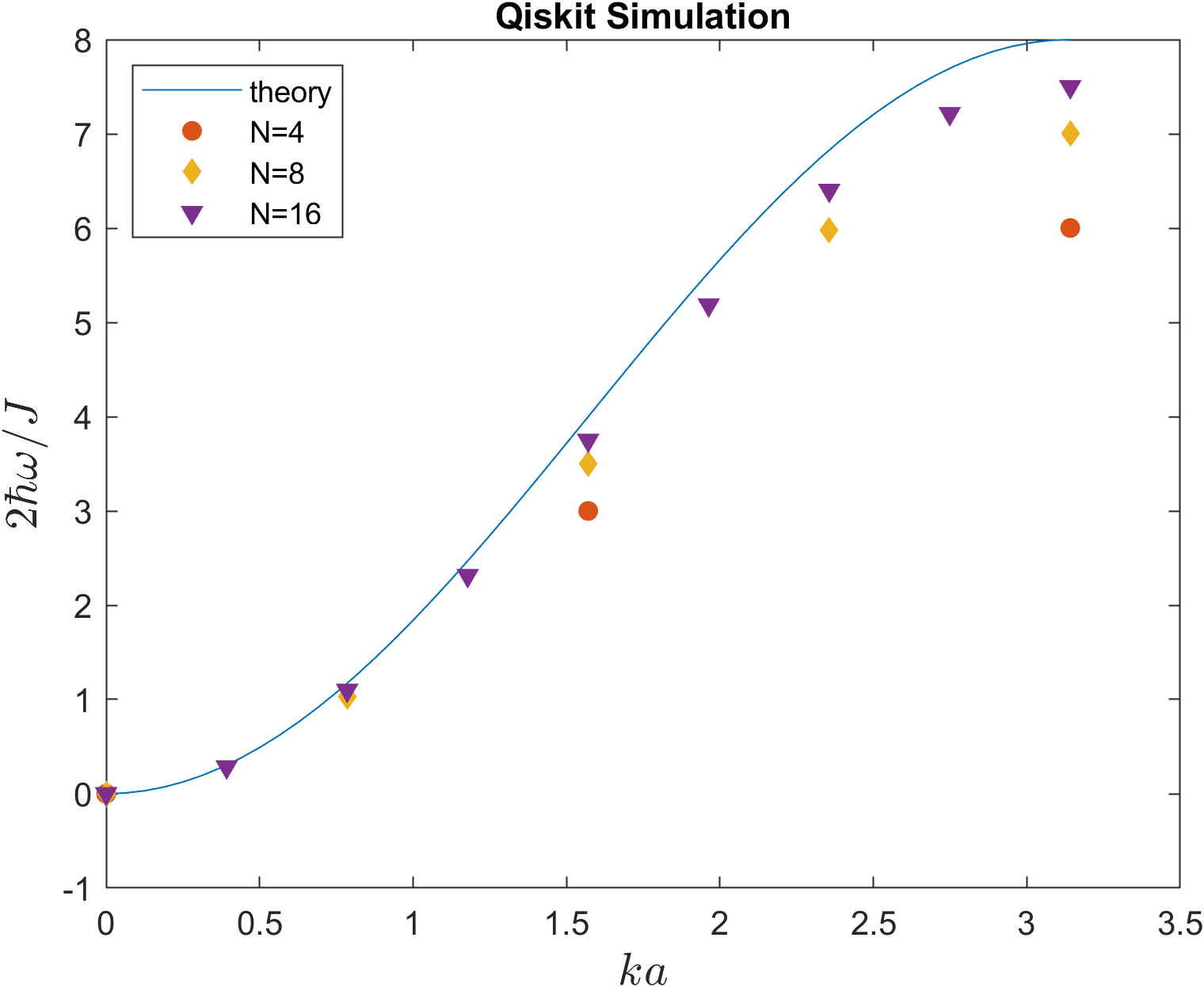}
    \caption{The dispersion relation computed with the ansatz (\ref{eq:ansatz}) appears to converge to the correct values for $N\rightarrow\infty$.}   \label{fig:convergence}
\end{figure}

\begin{figure}
    \centering
    \includegraphics[width=0.75\linewidth]{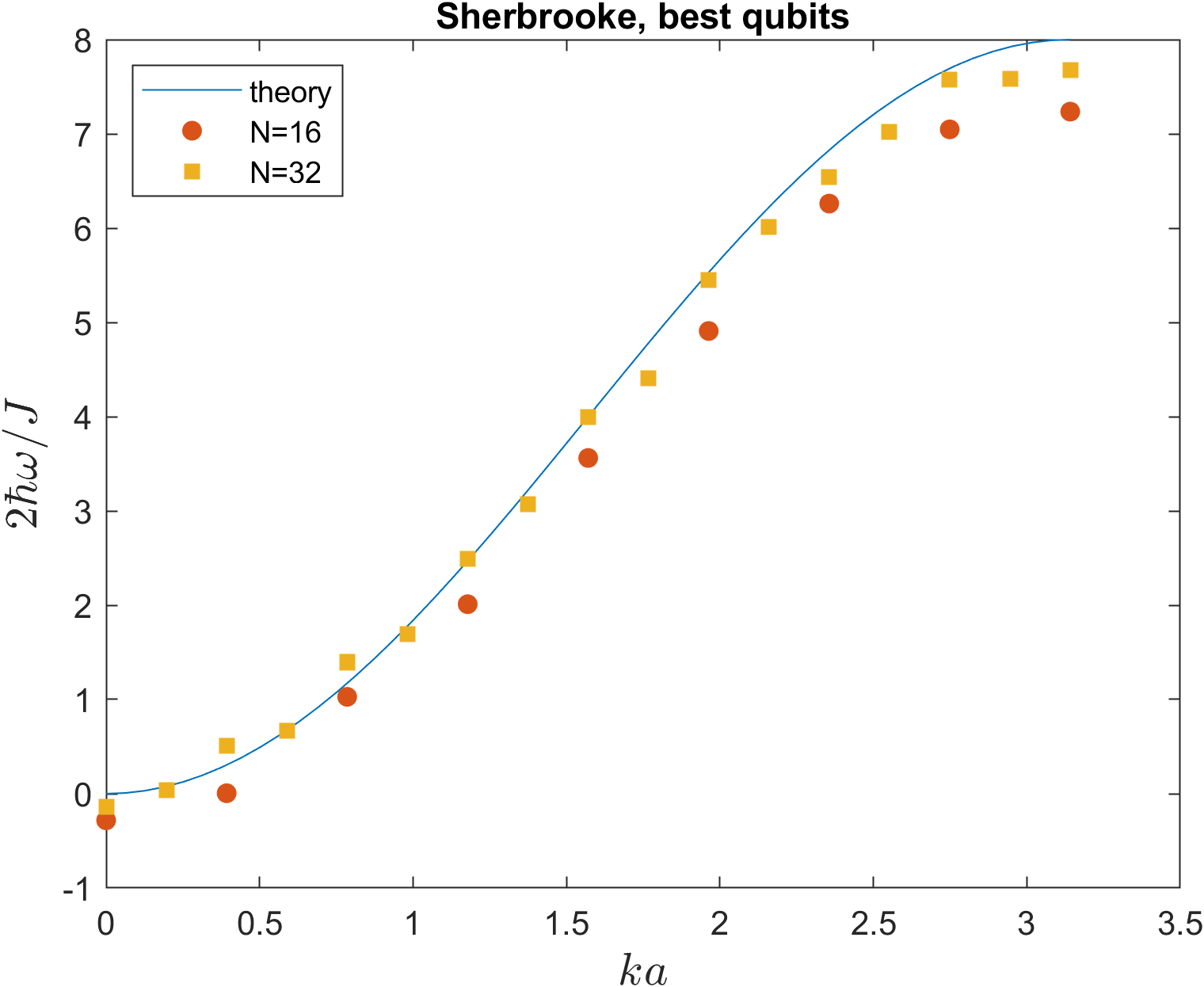}
    \caption{The dispersion relation computed with the ansatz on IBM Sherbrooke, using the qubits with the lowest single-gate error.}   \label{fig:Qprocessor}
\end{figure}

Calculations using the Qiskit simulator for the expected value of the
Hamiltonian (\ref{eq:Heisenberg_ham}) using the ansatz (\ref{eq:ansatz})
are shown in Fig. \ref{fig:convergence} along with the theoretical result (\ref{eq:theory}).
The calculations for $N=4,6,8$
show that the expected value of the energy using the wavefunction ansatz (\ref{eq:ansatz}) approaches the correct value as $N$ becomes large. This trend is further shown in Fig. \ref{fig:Qprocessor} for $N=16,32$ using the IBM Sherbrooke quantum processor. (Readout error mitigation was performed, using the twirled Pauli approach described by van den Berg \textit{et al.}~\cite{vandenBerg_trex_2022}.)

\begin{figure}
    \centering
    \includegraphics[width=0.9\linewidth]{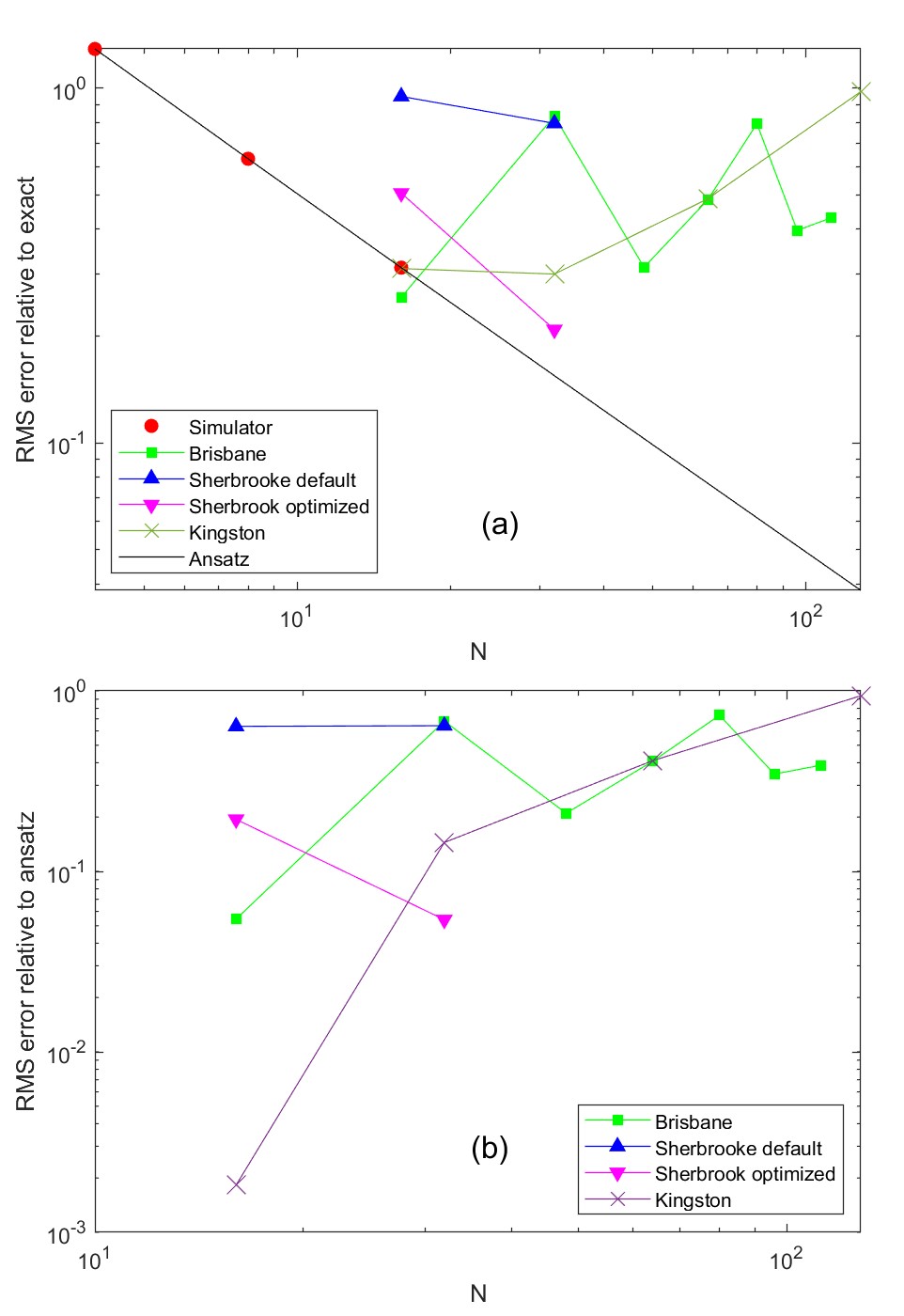}
    \caption{Dispersion relation RMS error for various processors. (a) Error relative to exact dispersion relation. The curve labeled ``Ansatz" is from Equation (\ref{eq:rms3}). (b) Magnitude of the error relative to the dispersion relation from the Ansatz}   \label{fig:error_combined}
\end{figure}

As shown in Appendix \ref{app:B}, the RMS difference between the exact solution (\ref{eq:theory}) and the Ansatz dispersion relation is given by
\begin{equation}
     \epsilon_{\text{RMS}}(N) =\frac{1}{N}\sqrt{\frac{8(3N+8)}{N+2}}.
     \label{eq:rms3}
\end{equation}
The actual RMS error obtained using various quantum processors is compared with this expression in Figure \ref{fig:error_combined}(a), and the RMS error relative to the Ansatz value is shown in Figure \ref{fig:error_combined}(b). 
The calculations on the IBM Brisbane processor used readout error mitigation and Zero Noise Extrapolation (ZNE) to minimize error, but used the default qubit assignments. Calculations on IBM Sherbrooke are shown for both the default qubit assignment, as well as the optimized assignment used in Figure \ref{fig:Qprocessor}. The calculations on IBM Kingston used the 
twirled Pauli approach mentioned earlier
\cite{vandenBerg_trex_2022}.
The accuracy is quite good for small $N$, but increases rapidly for larger $N$.

\section{Discussion}

Since simulators represent the total quantum state, simulation calculations become difficult beyond about 30 qubits. However, note that since the present calculation does not involve any two-qubit gates, it could be performed for an arbitrarily large system simply by repeatedly simulating a single qubit with varying rotation parameters and combining the result classically. Therefore, the only advantage of calculating larger systems on a quantum processor is classical parallelism.

The significance of using the quantum computing paradigm in this case is providing new insight into the relationship between the quantum and classical spin wave descriptions, rather than accelerated calculations. In particular, although the exact wave function for a linear chain of spins is an entangled state, it is surprising to discover that a product state with progressive $Z$-rotations gives the correct expected value for the dispersion relation as the number of qubits $N$ becomes large--even though the product state does not approach the correct entangled state in this limit.

The RMS error between the actual values computed with quantum processors and the theoretical values gives an indication of the single-qubit error across a processor. In particular, the larger the value of $N$, the more precise the $Y$ rotation must be. Since the error from single qubit rotations accumulates to limit the useful circuit depth, this also gives an indication of circuit depth limitations.
Thus this model may also be useful for evaluation of the error in quantum processors.

\section*{Acknowledgements}
We acknowledge the use of IBM Quantum via the IBM Quantum Innovation Center at NC State for this work. The views expressed are those of the authors and do not reflect the official policy or position of the IBM Quantum Innovation Center at NC State, IBM or the IBM Quantum team.
BNB was supported by the U.S. Department of Energy, Advanced Scientific Computing Research, under contract number DE-SC0025384.

\appendix
\section{Ansatz Dispersion Relation}
\label{app:A}
 The Hamiltonian (\ref{eq:Heisenberg_ham}) can be written as
 \begin{equation}
H=-\frac{J}{2}\sum_{n=0}^{N-1}\left( X_n X_{n+1}+Y_n Y_{n+1} + Z_n Z_{n+1}\right)\label{eq:Heisenberg_expand}
\end{equation}
Note that in the $n$th term, only the qubits $n$ and $n+1$ are affected by the Hamiltonian. 
Thus,
\begin{align}
 \braket{\psi_A|H|\psi_A}&=-\frac{J}{2}\sum_{n=0}^{N-1}\bra{\psi_n}\otimes\bra{\psi_{n+1}}\left( X_n X_{n+1}\right.\\\nonumber
 &\left.+Y_n Y_{n+1} + Z_n Z_{n+1}\right)\ket{\psi_n}\otimes\ket{\psi_{n+1}},
\end{align}
where
\begin{align}
 \ket{\psi_n} = \cos{\frac{\theta}{2}}\ket{0}+e^{inka}\sin{\frac{\theta}{2}}\ket{1}.
\end{align}
We now explicitly evaluate each term for a given value of $n$ as follows:
\begin{align}
\bra{\psi_n}&\otimes\bra{\psi_{n+1}}X_n X_{n+1}\ket{\psi_n}\otimes\ket{\psi_{n+1}} \nonumber \\
    &=\frac{\sin^2\theta}{2}\left(\cos{(ka)}+\cos{((2n+1)ka)}\right),
\end{align}
\begin{align}
\bra{\psi_n}&\otimes\bra{\psi_{n+1}}Y_n Y_{n+1}\ket{\psi_n}\otimes\ket{\psi_{n+1}} \nonumber \\
    &=\frac{\sin^2\theta}{2}\left(\cos{(ka)-\cos{((2n+1)ka)}}\right),
\end{align}
and
\begin{align}
\bra{\psi_n}\otimes\bra{\psi_{n+1}}Z_n Z_{n+1}\ket{\psi_n}\otimes\ket{\psi_{n+1}} =\cos^2\theta.
\end{align}
Adding the terms together and multiplying by $N$, since all summands are equal, gives
\begin{align}
    \braket{\psi_A|H|\psi_A}&=-\frac{JN}{2}(\sin^2\theta \cos{ka} +\cos^2\theta)\nonumber \\
    &=-\frac{JN}{2}\left(\sin^2\theta (\cos{ka} - 1) + 1\right).
    \label{eq:ham_sum}
\end{align}
But
\begin{align}
    \sin^2{\theta}&=4\sin^2\frac{\theta}{2}\cos^2{\frac{\theta}{2}}\nonumber \\
    &=4\sin^2{\frac{\theta}{2}}(1-\sin^2{\frac{\theta}{2}})\nonumber \\
    &=\frac{4}{N}\left(1-\frac{1}{N}\right),
    \label{eq:optangle2}
\end{align}
where use has been made of Eq. (\ref{eq:optangle}).
Substituting this result into (\ref{eq:ham_sum}) leads to
\begin{equation}
    \frac{2}{J}\braket{\psi_A|H|\psi_A}+N =4\left(1-\frac{1}{N}\right)(1-\cos{ka}).
    \label{eq:ansatzdisp}
\end{equation}
Taking the limit of large $N$ gives directly Eq. (\ref{eq:limit}).

\section{Ansatz RMS Error}
\label{app:B}
The squared difference between the Ansatz dispersion relation (\ref{eq:ansatzdisp}) and the correct expression (\ref{eq:theory}) is given by
\begin{equation}
    \Delta^2 = \frac{16}{N^2}(\cos{ka}-1)^2.
\end{equation}
It follows that the RMS error between the Ansatz and correct expressions is given by
\begin{equation}
    \epsilon_{\text{RMS}}(N) =\left[\frac{32}{N^2(N+2)}\sum_{n=0}^{N/2}\left(\cos{\frac{2\pi n}{N}}-1\right)^2\right]^{1/2}.
    \label{eq:rms1}
\end{equation}
With the assistance of Wolfram$|$Alpha \cite{wolframalpha2025}, the sum is found to be
\begin{equation}
    \sum_{n=0}^{N/2}\left(\cos{\frac{2\pi n}{N}}-1\right)^2=\frac{3N}{4}+2.
\end{equation}
Substituting this result into (\ref{eq:rms1}) and simplifying leads to
\begin{equation}
     \epsilon_{\text{RMS}}(N) =\frac{1}{N}\sqrt{\frac{8(3N+8)}{N+2}}.
     \label{eq:rms2}
\end{equation}

\bibliography{spinwaves}

\end{document}